\documentclass[aps,prb,preprint,superscriptaddress]{revtex4-2}

\usepackage{amsmath}
\usepackage{amssymb}
\usepackage{bbm}

\usepackage{braket}
\usepackage{units}

\usepackage{url}
\usepackage{hyperref}

\usepackage[pdftex]{graphicx}

\usepackage[T1]{fontenc}
\usepackage[utf8]{inputenc}

\renewcommand{\vec}[1]{\boldsymbol{#1}}
\newcommand{\mat}[1]{\mathsf{#1}}
\renewcommand{\Re}{\mathrm{Re}}
\renewcommand{\Im}{\mathrm{Im}}
\newcommand{\tr}{\mathrm{tr}}

\begin{document}

\title{A Unified Theoretical Framework for Photoemission and Its Inverse: Reciprocity, Spin, and Photon Polarization}

\author{Frank O. Schumann}
\affiliation{Max Planck Institute of Microstructure Physics, Weinberg 2, 06120 Halle, Germany}

\author{Jürgen Henk}
\affiliation{Martin Luther University Halle-Wittenberg, Quantum Theory of the Solid State, 06099 Halle, Germany}

\date{\today}

\begin{abstract}
We present a unified theoretical framework for spin- and angle-resolved photoemission spectroscopy (SARPES) and its inverse process (SARIPES), treating both on an equal footing. The formulation is based on a response tensor that encodes the reciprocity between the two processes. We show that polarization tomography within SARPES calculations provides a systematic route for reconstructing the tensor elements. The number of independent components is significantly reduced by symmetry constraints, which in turn allows for explicit expressions describing spin-reversal processes and intensity modulations. The formalism is illustrated for the W(110) surface. The resulting framework provides a symmetry-adapted description of spin-dependent optical transitions at surfaces and enables the direct prediction of SARIPES intensities from SARPES calculations.
\end{abstract}

\maketitle

\section{Introduction}
Spin- and angle-resolved photoemission spectroscopy (SARPES) and spin- and angle-resolved inverse photoemission spectroscopy (SARIPES) probe complementary aspects of the same optical transition processes at surfaces~\cite{Huefner2003,Schattke2004,Suga2021}. In SARPES, an incident photon excites an electron from an occupied state of the solid into the vacuum, where its momentum and spin polarization can be analyzed~\cite{Dil2009}. In SARIPES~\cite{Dose1985}, the time-reversed process occurs: a beam of spin-polarized electrons impinges on the surface, occupies unoccupied electronic states of the solid, and emits photons that are detected outside the sample~\cite{Pendry1980,Pendry1981,Borstel1988}. Although the experimental observables are different in the two methods, both processes are governed by the same dipole transition matrix elements that connect electronic states in the solid with continuum states in the vacuum.

This relationship implies a fundamental reciprocity between SARPES and SARIPES\@. In SARPES, the polarization state of the incident photon field is typically controlled, and the emitted electron intensity and spin polarization are measured. In SARIPES, however, the situation is reversed: the spin polarization of the incoming electron beam is prepared, while the intensity of the emitted photons is detected. Establishing an explicit connection between these two experimental configurations therefore provides a natural framework for translating information obtained in SARPES into predictions for SARIPES, and vice versa.

A convenient formulation of this reciprocity is provided by a response tensor that links the polarization state of the photon field to the spin polarization of the electrons~\cite{Donath1991}. The tensor is constructed from response functions and contains complete information on spin-dependent optical transitions. The measured SARPES or SARIPES intensity follows from its combination with the photon polarization and the spin degrees of freedom, which enter in complementary roles in the two processes. In this way, the response tensor provides a unified description of both techniques and renders their reciprocity transparent.

In this Paper we present a theoretical investigation that follows the conceptual framework outlined above. We first introduce the general formalism for spin- and polarization-resolved transitions and formulate the optical matrix elements within the Stokes description of photon polarization. On this basis, we construct a response tensor that links photon polarization and electron spin polarization, thereby making the reciprocity between SARPES and SARIPES explicit. We then analyze the symmetry properties of this tensor and show how spatial symmetries constrain the number of independent tensor elements. Within this framework, we demonstrate by numerical calculations how polarization tomography enables the construction of the response tensor. Eventually, we apply the formalism to SARPES and SARIPES from W(110), showing explicitly how SARIPES intensities can be predicted from SARPES calculations. The aim of this study is not to reproduce published experimental data, but to outline the proposed framework and demonstrate its feasibility through proof-of-principle calculations.

The Paper is organized as follows. In Section~\ref{sec:reciprocity} we discuss the reciprocity of SARPES and SARIPES for nonmagnetic systems. The SARPES intensities, introduced in Section~\ref{sec:SARPES-intensities}, are reformulated in terms of a response tensor in Section~\ref{sec:response-tensor}, which in turn provides direct access to the SARIPES intensities. In Section~\ref{sec:example} the formalism is applied to the case of a W(110) surface. Additional technical details are provided in the appendices, which include further discussion of the response tensor (\ref{sec:derivation-response-tensor} as well as \ref{sec:strategy-response-tensor}) and of the numerical calculations (\ref{sec:numerical}). Supporting data are given in Appendix~\ref{sec:supporting}.

\section{Reciprocity of SARPES and SARIPES}
\label{sec:reciprocity}
The relation between the transition matrix elements of conventional and inverse photoemission can be understood from the reciprocity of electron scattering states~\cite{Pendry1976,Braun1996,Minar2011}. In the one-electron picture the photoemission intensity is determined by the dipole matrix element
\begin{align*}
    D_{\mathrm{vc}} = \braket{\psi_{\mathrm{v}} | \hat{H}_{\mathrm{int}} | \psi_{\mathrm{c}}},
\end{align*}
where $\ket{\psi_{\mathrm{c}}}$ is the electron state in the crystal and $\ket{\psi_{\mathrm{v}}}$ the state describing the emitted electron in vacuum. The interaction Hamiltonian $\hat{H}_{\mathrm{int}}$ corresponds to the dipolar coupling to the electromagnetic field~\cite{Eberhardt1980}.

The state $\ket{\psi_{\mathrm{v}}}$ is not a simple plane wave but a scattering state that propagates through the crystal potential and emerges into the vacuum~\cite{Feibelman1974}. Following the reciprocity argument introduced by Pendry~\cite{Pendry1976}, this state can be replaced by a time-reversed LEED state~\cite{Feibelman1974} (low-energy electron diffraction). In other words, instead of describing an electron leaving the crystal one may equivalently consider the time-reversed problem of an electron incident on the surface from the vacuum.

Denoting by $\ket{\psi_{\mathrm{LEED}}^{(+)}}$ the LEED state corresponding to an electron incident from vacuum, the photoemission final state $\ket{\psi_{\mathrm{v}}}$ becomes $\ket{\psi_{\mathrm{LEED}}^{(-)}}$, where the superscript $(-)$ indicates the time-reversed scattering boundary condition. The photoemission matrix element can therefore be expressed as
\begin{align}
    D_{\mathrm{vc}} = \braket{\psi_{\mathrm{LEED}}^{(-)} | \hat{H}_{\mathrm{int}} | \psi_{\mathrm{c}}}.
    \label{eq:MEvc}
\end{align}

\begin{figure*}
    \centering
    \includegraphics[width = 0.9 \textwidth]{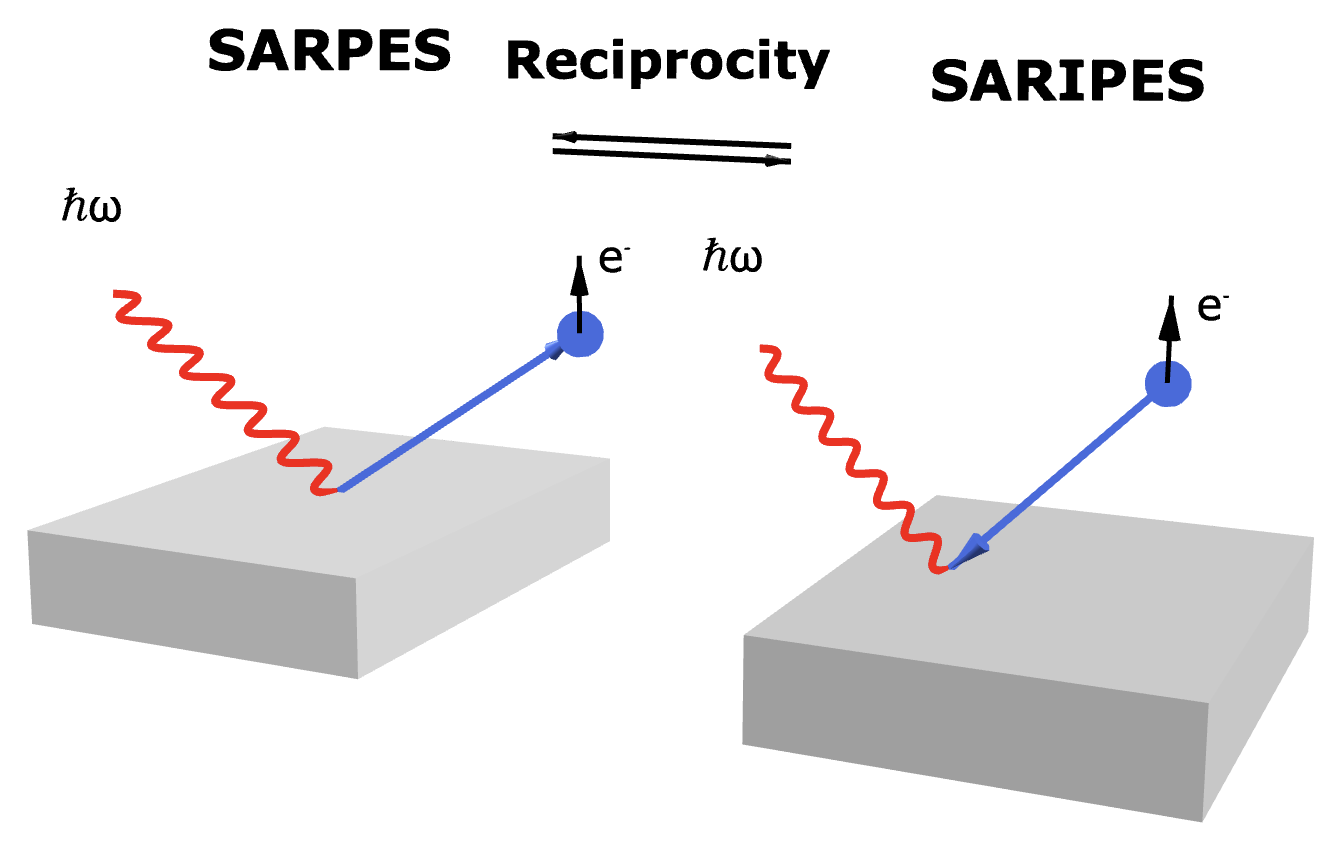}
    \caption{Reciprocity of spin- and angle-resolved photoemission spectroscopy (SARPES, left) and its inverse process (SARIPES, right). In SARPES, an electromagnetic wave with photon energy~$\hbar\omega$ (red sinosoidal curve) impinges on the sample (gray block), leading to the emission of a beam of spin-polarized electrons ($e^{-}$, blue sphere). In SARIPES, an incident beam of spin-polarized electrons gives rise to the emission of electromagnetic radiation.}
    \label{fig:reciprocity}
\end{figure*}

In inverse photoemission, the incoming electron is described directly by the LEED state $\ket{\psi_{\mathrm{LEED}}^{(+)}}$. The corresponding transition matrix element reads
\begin{align*}
    D_{\mathrm{cv}} = \braket{\psi_{\mathrm{c}} | \hat{H}_{\mathrm{int}} | \psi_{\mathrm{LEED}}^{(+)}}.
\end{align*}
Because the Hamiltonian is Hermitian~\footnote{This is not the case when self-energy effects are included.} and the scattering states are
related by time reversal, the two transition probabilities are equal
up to the reversal of momentum $\hbar \vec{k}$ and spin $\sigma$,
\begin{align}
    \left| D^{(\sigma)}_{\mathrm{SARPES}}(\vec{k}, \omega) \right|^{2} & = \left| D^{(-\sigma)}_{\mathrm{SARIPES}}(-\vec{k}, \omega) \right|^{2}.
    \label{eq:tme}
\end{align}
This reciprocity expresses the fact that the probability for an electron to leave the crystal and reach the SARPES detector is equal to the probability for a time-reversed electron trajectory entering the crystal from the detector direction (Figure~\ref{fig:reciprocity}). Consequently, the same scattering amplitudes determine the matrix elements in both conventional and inverse photoemission.

We would like to stress that, while SARPES and SARIPES are both governed by the same transition matrix elements, Equation~\eqref{eq:tme}, the origin of their spin sensitivity is different. In SARPES, the measured spin polarization emerges during the photoemission process itself. In contrast, in SARIPES the spin sensitivity is externally imposed by the incoming spin-polarized electron beam, which may be viewed as a projection filter onto the spin-resolved spectral density of the sample; the corresponding response is reflected in the emitted electromagnetic field. Although both techniques probe the same underlying spin–orbit--coupled electronic structure and therefore often show good agreement with the spin-resolved spectral density, this correspondence should not be interpreted as a direct one-to-one mapping of spin polarization between the two experiments.

\section{Spin- and angle-resolved conventional photoemission}
\label{sec:SARPES-intensities}
Consider a monochromatic electromagnetic plane wave with photon energy $\hbar \omega$ incident onto the surface of a sample under polar angle $\vartheta$ (with respect to the surface normal $z$) and azimuth $\varphi$ (measured from the $x$ axis in the surface plane). The propagation direction of the light is therefore
\begin{align*}
    \vec{q}(\vartheta, \varphi) & =
    \begin{pmatrix}
    \sin\vartheta \cos\varphi, 
    \sin\vartheta \sin\varphi,
    -\cos\vartheta
    \end{pmatrix}^{\mathrm{t}}.
\end{align*}

In this Paper we utilize three descriptions of the electromagnetic wave: the $(s,p)$ basis, a Cartesian basis, and Stokes parameters. In the $(s,p)$ polarization basis, the electric field of the incident light is written as
\begin{align}
    \vec{E}(\vartheta, \varphi) & = E_{s}\,\vec{e}_{s}(\varphi) + E_{p}\,\vec{e}_{p}(\vartheta, \varphi),
    \label{eq:e-field-cartesian-sp}
\end{align}
where $E_{s}$ and $E_{p}$ are the complex amplitudes of the $s$- and $p$-polarized components. The $s$-polarization vector 
\begin{align*}
    \vec{e}_{s}(\varphi) & =
    \begin{pmatrix}
    -\sin\varphi, \cos\varphi, 0
    \end{pmatrix}^{\mathrm{t}}
    \end{align*}
is perpendicular to the incidence plane, while the $p$-polarization vector 
\begin{align*}
    \vec{e}_{p}(\vartheta, \varphi) & =
    \begin{pmatrix}
    \cos\vartheta \cos\varphi, \cos\vartheta \sin\varphi, \sin\vartheta
    \end{pmatrix}^{\mathrm{t}}
    \end{align*}
lies in the incidence plane; $\vec{e}_{s}$, $\vec{e}_{p}$, and $\vec{q}$ form an orthogonal tripod.  The Cartesian components of the electric field are then given by Equation~\eqref{eq:e-field-cartesian-sp}.

In addition, the electric field can be expressed in terms of the Stokes vector $\vec{S} = (S_{0}, S_{1}, S_{2}, S_{3})^{\mathrm{t}}$, with components
\begin{subequations}
\label{eq:Stokes}
\begin{align}
    S_{0} & = |E_{s}|^{2} + |E_{p}|^{2}, \quad & \text{total intensity}\\
    S_{1} & = |E_{s}|^{2} - |E_{p}|^{2}, \quad & \text{linear } (s,p)\\
    S_{2} & = 2\,\mathrm{Re}(E_{s} E_{p}^{\star}), \quad & \text{linear ±45° w.r.t. } (s,p)\\
    S_{3} & = 2\,\mathrm{Im}(E_{s} E_{p}^{\star}). \quad & \text{circular  ±45° w.r.t. } (s,p)
\end{align}
\end{subequations}

In a SARPES experiment, one detects photoelectrons spin-polarized along the Cartesian axis $\sigma$ and with momentum $\hbar \vec{k}$ for the given photon energy $\hbar\omega$ (left in Figure~\ref{fig:reciprocity}). The spin-resolved intensities depend to lowest order quadratically on the electric field of the incident radiation; hence
\begin{align}
    I^{(\pm \sigma)}(\vec{\pi}) & = \sum_{i, j} E_{i} \, M^{(\pm \sigma)}_{ij}(\vec{\pi}) E_{j}^{\star},
    \label{eq:intensity}
\end{align}
in which $i$ and $j$ run over $x$, $y$, and $z$ in the Cartesian basis [Equation~\eqref{eq:e-field-cartesian-sp}] or over $s$ and $p$ in the $(s, p)$ basis; $\vec{\pi} = (\vec{k}, \omega, \vec{q})$ comprises the parameters defining the chosen setup. $M^{(\pm \sigma)}_{ij}$ are spin-dependent response functions for spin direction $\sigma \in \{ x, y, z \}$ [$\pm \sigma$ is short for parallel/anti-parallel to the $\sigma$ axis, respectively; these functions contain transition matrix elements, such as defined in Equation~\eqref{eq:MEvc}]. The reality of $I^{(\pm \sigma)}$ implies $M^{(\pm \sigma)}_{ij} = M^{(\pm \sigma)\star}_{ji}$. In terms of the Stokes parameters the intensity reads
\begin{align}
    I^{(\pm \sigma)}(\vec{\pi}) & = 
    \sum_{i = 0}^{3} a_{i}^{(\pm \sigma)}(\vec{\pi}) \, S_{i},
    \label{eq:intensity-Stokes-basis}
\end{align}
with
\begin{subequations}
\begin{align}
    a_{0}^{(\pm \sigma)} & = \frac{1}{2} \left( M_{ss}^{(\pm \sigma)} + M_{pp}^{(\pm \sigma)} \right),\\
    a_{1}^{(\pm \sigma)} & = \frac{1}{2} \left( M_{ss}^{(\pm \sigma)} - M_{pp}^{(\pm \sigma)} \right), \\
    a_{2}^{(\pm \sigma)} & = \Re\, M_{sp}^{(\pm \sigma)}, \\
    a_{3}^{(\pm \sigma)} & = -\Im\, M_{sp}^{(\pm \sigma)}.
\end{align}    
    \label{eq:coeffs-Stokes-basis}
\end{subequations}

\section{Spin–Stokes response tensor}
\label{sec:response-tensor}
The key quantity in SARPES is the spin polarization vector of the outgoing photoelectrons. For the purpose of this Paper we extend this Cartesian $3$-vector by a fourth component, $\vec{P} = (P_{0}, P_{x}, P_{y}, P_{z})^{\mathrm{t}}$, with
\begin{subequations}
\begin{align}
    P_{0}      & = 1, \\
    P_{\sigma}(\vec{\pi}) & = \frac{I^{(+\sigma)}(\vec{\pi}) - I^{(-\sigma)}(\vec{\pi})}{I^{(+\sigma)}(\vec{\pi}) + I^{(-\sigma)}(\vec{\pi})}, \quad \sigma = x, y, z \text{ or } 1, 2, 3. \label{eq:Pdef}
\end{align}    
\end{subequations}
The component $P_{0}$ accounts for the spin-independent (spin-averaged) part of the photoemission intensity. $P_{\sigma}(\vec{\pi})$ quantifies the usual spin polarization~\cite{Kessler1985}, given by the spin-resolved intensities $I^{(\pm \sigma)}(\vec{\pi})$ defined in the previous Section, with respect to the Cartesian spin axis~$\sigma$; it satisfies $|P_{\sigma}| \le 1$. 

The intensities depend linearly on the Stokes parameters [Equation~\eqref{eq:intensity-Stokes-basis}] and, within the spin-resolved decomposition, can be written as
\begin{align*}
    I^{(\pm \sigma)}(\vec{\pi})
    & = \frac{1}{2} \left[ P_{0} \pm P_{\sigma}(\vec{\pi}) \right] I(\vec{\pi}) .
\end{align*}
In this representation, the spin-resolved signals are linear combinations of the spin-polarization components, with $I(\vec{\pi})$ denoting the spin-averaged intensity. This suggests to introduce a response tensor $\mathcal{R}(\vec{\pi})$, which allows to express the intensity in a bilinear form,
\begin{align}
    I(\vec{\pi}; \vec{S}, \vec{P}) & = \sum_{\alpha, \beta = 0}^{3} S_{\alpha} \, \mathcal{R}_{\alpha\beta}(\vec{\pi}) \, P_{\beta}.
    \label{eq:bilinear}
\end{align}
$\mathcal{R}$ acts as a kernel connecting the two polarization spaces: it gives the intensity for each combination of Stokes parameters $S_{\alpha}$ and spin polarization components $P_{\beta}$, combining `cause and effect' or `input and output'. Its explicit form is derived in Appendix~\ref{sec:derivation-response-tensor}.

Considering now the reciprocity of SARPES and SARIPES (Section~\ref{sec:reciprocity}) the interpretation of $\vec{S}$ and $\vec{P}$ in Equation~\eqref{eq:bilinear} becomes context-dependent: depending on the experimental situation (SARPES or SARIPES), either vector can play the role of `source' or `response'. In a SARPES context (Stokes parameters $S_{\alpha}$ as source), the contraction
\begin{align}
    \tilde{P}_{\beta}(\vec{\pi}; \vec{S}) & = \sum_{\alpha = 0}^{3} S_{\alpha} \mathcal{R}_{\alpha \beta}(\vec{\pi}), \quad \beta = 0, \ldots, 3,
    \label{eq:spinpol-contraction}
\end{align}
yields as a response an effective electron spin-polarization 4-vector $\tilde{\vec{P}}$. Rewriting the intensity as
\begin{align*}
    I(\vec{\pi}; \vec{S}, \vec{P})
    = \sum_{\beta = 0}^{3} \tilde{P}_{\beta}(\vec{\pi}; \vec{S}) \, P_{\beta},
\end{align*}
makes the dependence on the detector setting $\vec{P}$ explicit. The vector $\tilde{\vec{P}}(\vec{\pi}; \vec{S})$ can be interpreted as the induced polarization produced by the transition process for a given source polarization $\vec{S}$. The measured intensity is then obtained by projecting this induced polarization onto $\vec{P}$, which plays the role of an analyzer (detector) polarization.

Equation~\eqref{eq:bilinear} admits an analogous interpretation in the SARIPES context, where the electron spin polarization $\vec{P}$ serves
as the source variable. The contraction
\begin{align}
    \tilde{S}_{\alpha}(\vec{\pi}; \vec{P}) 
    = \sum_{\beta = 0}^{3} \mathcal{R}_{\alpha\beta}(\vec{\pi}) \, P_{\beta},
    \quad \alpha = 0, \ldots, 3,
    \label{eq:Stokes-inverse}
\end{align}
defines an effective Stokes 4-vector $\tilde{\vec{S}}(\vec{\pi}; \vec{P})$. It reproduces the intensity as
\begin{align*}
    I(\vec{\pi}; \vec{S}, \vec{P}) 
    = \sum_{\alpha = 0}^{3} S_{\alpha} \, \tilde{S}_{\alpha}(\vec{\pi}; \vec{P}).
\end{align*}
The vector $\tilde{\vec{S}}$ can be interpreted as the induced Stokes vector of the emitted radiation generated by a given electron spin polarization $\vec{P}$. The measured intensity is then obtained by
projecting this induced Stokes vector onto $\vec{S}$, which represents
the analyzer setting.

Once the effective Stokes parameters $\tilde{S}_{\alpha}$ are known, the complex electric-field components in the $(s, p)$ basis follow from Equation~\eqref{eq:Stokes}:
\begin{subequations}
\begin{align}
    \tilde{E}_{s} & = \sqrt{\frac{\tilde{S}_{0} + \tilde{S}_{1}}{2}}, \\
    \tilde{E}_{p} & = \sqrt{\frac{\tilde{S}_{0} - \tilde{S}_{1}}{2}} \, e^{-\mathrm{i} \tilde{\delta}},
\end{align}    
\label{eq:es-ep}
\end{subequations}
with
\begin{align}
    \tan\tilde{\delta} & = \frac{\tilde{S}_{3}}{\tilde{S}_{2}}
    \label{eq:es-ep-delta}
\end{align}
(only an irrelevant global phase remains undetermined).

In general, it is not guaranteed that $\tilde{\vec{P}}$  and $\tilde{\vec{S}}$ satisfy the positivity constraints for polarization vectors, e.\,g. $\tilde{P}_{x}^{2} + \tilde{P}_{y}^{2} + \tilde{P}_{z}^{2} \leq 1$~\footnote{The mapping between electron spin polarization and emitted photon polarization is naturally described within the quantum operations formalism, where physical transformations are represented by completely positive maps in Kraus form \cite{NielsenChuang2000}. Similar positivity constraints arise in classical polarization optics, where Mueller matrices must map the solid light cone of Stokes vectors into itself \cite{Simon2010Mueller,GamelJames2011}.}. A sufficient condition for positivity is that the response tensor $\mathcal{R}$ defines a completely positive map, which is the case when $\mathcal{R}$ is obtained from Fermi’s golden rule.

The bilinear form~\eqref{eq:bilinear} is central to this study, since the SARPES--SARIPES reciprocity emerges naturally. The elements $\mathcal{R}_{\alpha\beta}$ of the response tensor quantify the coupling between the components $S_{\alpha}$ and $P_{\beta}$. Depending on the context, this coupling admits two complementary interpretations: in SARPES, the source $\vec{S}$ induces an electron spin polarization $\tilde{\vec{P}}$ according to Equation~\eqref{eq:spinpol-contraction}, whereas in SARIPES, the source $\vec{P}$ induces a Stokes vector $\tilde{\vec{S}}$ according to Equation~\eqref{eq:Stokes-inverse}. Within these interpretations, $I(\vec{\pi}; \vec{S}, \vec{P})$ is not regarded as a directly measurable observable, but acts as a generating quantity  (Appendix~\ref{sec:strategy-response-tensor}) whose operational meaning is revealed only through projection onto the corresponding analyzer (detector) setting.

In the Appendices, we derive explicit expressions for the response tensor (\ref{sec:derivation-response-tensor}) and suggest a strategy how derive its elements from SARPES calculations (\ref{sec:strategy-response-tensor}). It is important to note that the response tensor $\mathcal{R}$ is specific to the setup $(\vec{\pi})$ used in these `calibrating' calculations, since it implicitly depends not only on the orientation of the crystal but also on the emission direction $\vec{k}$, the incidence direction $\vec{q}$, and the photon energy $\hbar\omega$. Consequently, the SARPES-calibrated tensor can be used to describe the reciprocal SARIPES process only for the corresponding time-reversed (SARIPES) scenario. Within this fixed setup, however, $\mathcal{R}$ provides a complete mapping between spin polarization of the electron beam and the polarization state of the radiation.

\section{Application: photoemission in a mirror plane}
\label{sec:example}
Having introduced the main idea of the response-tensor approach, we now demonstrate its capabilities through analytical and numerical calculations. For this purpose, we consider a surface with a mirror plane (here: the $xz$ plane, with $z$ the surface normal).

\subsection{Conventional photoemission}
\label{sec:conventional-PE}
The photoelectrons are detected off-normally within the $xz$ mirror plane. The relevant symmetry operation is therefore the reflection $m_{xz}: (x,y,z) \to (x,-y,z)$ at the $xz$ plane, under which the electric-field components transform as a polar vector,
\begin{align*}
    (E_{x},E_{y},E_{z}) & \to (E_{x},-E_{y},E_{z}),
\end{align*}
whereas the spin polarization transforms as an axial vector,
\begin{align*}
    (P_{x},P_{y},P_{z}) & \to (-P_{x},P_{y},-P_{z}).
\end{align*}
$E_{x}$, $P_{y}$, and $E_{z}$ are therefore even with respect to the mirror plane, whereas $P_{x}$, $E_{y}$, and $P_{z}$ are odd (Table~\ref{tab:symmetry-2mm}).

For radiation impinging  within the mirror plane the $(s, p)$ polarization basis is [$\varphi = \unit[0]{^{\circ}}$ in Equation~\eqref{eq:e-field-cartesian-sp}]
\begin{align*}
    \vec{E}_{s} & = E_{s} \, (0, 1, 0)^{\mathrm{t}},\\
    \vec{E}_{p} & = E_{p} \, (\cos\vartheta, 0, \sin\vartheta)^{\mathrm{t}},
\end{align*}
and the respective Stokes parameters transform under the mirror operation either as even ($S_{0}$ and $S_{1}$) or odd ($S_{2}$ and $S_{3}$).

\begin{table}
    \centering
    \begin{tabular}{c|ccc|c}
	\hline \hline
	$\mathbbm{1}$ &  $(E_{x},  E_{y}, E_{z})$ & $(E_{s},  E_{p})$ & $(S_{0}, S_{1}, S_{2}, S_{3})$ &  $(P_{x},  P_{y}, P_{z})$  \\
	$m_{xz}$      &  $(E_{x}, -E_{y}, E_{z})$ & $(-E_{s}, E_{p})$  & $(S_{0}, S_{1}, -S_{2}, -S_{3})$  &  $(-P_{x},  P_{y}, -P_{z})$ \\
	\hline \hline
    \end{tabular}
    \caption{Effect of the mirror operation $m_{xz}$ on the electric field $\vec{E}$ (represented in Cartesian coordinates, in the $(s,p)$ basis, and by Stokes parameters) and on the spin-polarization 3-vector $\vec{P}$. $\mathbbm{1}$ is the identity operation.}
    \label{tab:symmetry-2mm}
\end{table}

Since the SARPES intensity is invariant under the mirror symmetry~\cite{Hermanson1977} each term $S_{\alpha} P_{\beta}$ in the bilinear expression~\eqref{eq:bilinear} must be even. This means that even (odd) $S_{\alpha}$ can be combined with even (odd) $P_{\beta}$; all other tensor elements vanish by symmetry. The response tensor therefore takes the form
\begin{align}
    \mathcal{R} & =
    \begin{pmatrix}
    \mathcal{R}_{00} & 0 & \mathcal{R}_{0y} & 0\\
    \mathcal{R}_{10} & 0 & \mathcal{R}_{1y} & 0\\
    0 & \mathcal{R}_{2x} & 0 & \mathcal{R}_{2z}\\
    0 & \mathcal{R}_{3x} & 0 & \mathcal{R}_{3z}
\end{pmatrix}.
\label{eq:R-in-mirror-plane}
\end{align}

We now discuss briefly the contraction~\eqref{eq:spinpol-contraction} for the effective spin polarization $\tilde{\vec{P}}$, which in the present case provides
\begin{align*}
   \tilde{P}_{0} & = S_{0} \mathcal{R}_{00} + S_{1} \mathcal{R}_{10},
   \\
   \tilde{P}_{x} & = S_{2} \mathcal{R}_{2x} + S_{3} \mathcal{R}_{3x},
   \\
   \tilde{P}_{y} & = S_{0} \mathcal{R}_{0y} + S_{1} \mathcal{R}_{1y},
   \\
   \tilde{P}_{z} & = S_{2} \mathcal{R}_{2z} + S_{3} \mathcal{R}_{3z}.
\end{align*}
For incident s-polarized light (with Stokes parameters $S_{0} = S_{1} = |E|^{2}$, $S_{2} = S_{3} = 0$) these equations reduce to (see also Appendix~\ref{sec:derivation-response-tensor})
\begin{align*}
   \tilde{P}_{0} & = \left(M_{ss}^{(+y)} + M_{ss}^{(-y)} \right) |E|^{2},
   \\
   \tilde{P}_{y} & = \left( M_{ss}^{(+y)} - M_{ss}^{(-y)} \right) |E|^{2},
\end{align*}
and $\tilde{P}_{x} = \tilde{P}_{z} = 0$. $\tilde{P}_{0}$ is the total (spin-independent) intensity, while $\tilde{P}_{y} / \tilde{P}_{0}$ is the conventional $y$-spin polarization. Likewise, for p-polarized light ($S_{0} = |E|^{2}$, $S_{1} = -|E|^{2}$, $S_{2} = S_{3} = 0$), we have
\begin{align*}
   \tilde{P}_{0} & = \left(M_{pp}^{(+y)} + M_{pp}^{(-y)} \right) |E|^{2},
   \\
   \tilde{P}_{y} & = \left( M_{pp}^{(+y)} - M_{pp}^{(-y)} \right) |E|^{2},
\end{align*}
and $\tilde{P}_{x} = \tilde{P}_{z} = 0$, which is interpreted readily as for s-polarized light. Hence, we reveal the spin-polarization effect predicted and confirmed earlier~\cite{Tamura1991,Irmer1992}.

Considering left- and right-handed circular polarized light ($S_{0} = |E|^{2}$, $S_{1} = 0$, $S_{2} = 0$, $S_{3} = \pm |E|^{2}$) we find a superposition of the results for s- and p-polarized (in particular for $\tilde{P}_{0}$ and $\tilde{P}_{y}$), but in addition nonzero $x$- and $z$-components of $\tilde{\vec{P}}$ occur:
\begin{align*}
   \tilde{P}_{0} & = \frac{1}{2} |E|^{2} \left( 
   M_{ss}^{(+y)} + M_{ss}^{(-y)} + M_{pp}^{(+y)} + M_{pp}^{(-y)}
   \right),
   \\
   \tilde{P}_{x} & = \pm |E|^{2} \left( -\Im M_{sp}^{(+x)} + \Im M_{sp}^{(-x)}\right),
   \\
   \tilde{P}_{y} & = \frac{1}{2} |E|^{2} \left( M_{ss}^{(+y)} - M_{ss}^{(-y)} + M_{pp}^{(+y)} - M_{pp}^{(-y)} \right) ,
   \\
   \tilde{P}_{z} & = \pm |E|^{2} \left( -\Im M_{sp}^{(+z)} + \Im M_{sp}^{(-z)}\right).
\end{align*}
$\tilde{P}_{y} / \tilde{P}_{0}$ is independent of the helicity and in line with the respective expressions for s- or p-polarized light, since it comprises the same matrix elements. In contrast, the components within the scattering plane $\tilde{P}_{x} / \tilde{P}_{0}$ and  $\tilde{P}_{z} / \tilde{P}_{0}$ change sign upon helicity reversal, revealing optical orientation as their origin. Optical orientation describes the transfer of angular momentum from photons to electrons in a solid, whereby the helicity of circularly or elliptically polarized light induces a preferential spin polarization of the excited carriers~\cite{Meier1984}.

\subsection{Inverse photoemission}
For the reciprocal SARIPES process the effective Stokes parameters of the emitted radiation follow from contraction~\eqref{eq:Stokes-inverse}:
\begin{subequations}
\begin{align}
    \tilde{S}_{0} &= \mathcal{R}_{00} + \mathcal{R}_{0y} \, P_{y},\\
    \tilde{S}_{1} &= \mathcal{R}_{10} + \mathcal{R}_{1y} \, P_{y},\\
    \tilde{S}_{2} &= \mathcal{R}_{2x} \, P_{x} + \mathcal{R}_{2z} \, P_{z},\\
    \tilde{S}_{3} &= \mathcal{R}_{3x} \, P_{x} + \mathcal{R}_{3z} \, P_{z}
    \end{align}
    \label{eq:S-of-P}
\end{subequations}
(recall that $P_{0} = 1$).

\subsubsection{Incident spin polarization perpendicular to the mirror plane}
\label{sec:spin-perp-plane}
If the incident electron beam is polarized along the $y$ direction, $\vec{P} =(1,0,P_{y},0)^{\mathrm{t}}$, one obtains $\tilde{S}_{2} = \tilde{S}_{3} = 0$ [Equation~\eqref{eq:S-of-P}]. A spin flip $P_{y} \to -P_{y}$ results in a change of the respective Stokes parameters $\delta \tilde{S}_{i} = \tilde{S}_{i}(+P_{y}) - \tilde{S}_{i}(-P_{y})$. The relevant elements of response tensor read (Appendix~\ref{sec:derivation-response-tensor}) 
\begin{align*}
    \mathcal{R}_{0 0} &= \frac{1}{2} \left( M_{ss}^{(+y)} + M_{ss}^{(-y)} + M_{pp}^{(+y)} + M_{pp}^{(-y)} \right),\\
    \mathcal{R}_{0 y} &= \frac{1}{2} \left(M_{ss}^{(+y)} - M_{ss}^{(-y)} + M_{pp}^{(+y)} - M_{pp}^{(-y)} \right),
    \\
    \mathcal{R}_{1 0} &= \frac{1}{2} \left(M_{ss}^{(+y)} + M_{ss}^{(-y)} - M_{pp}^{(+y)} - M_{pp}^{(-y)}\right),\\
    \mathcal{R}_{1 y} &= \frac{1}{2} \left(M_{ss}^{(+y)} - M_{ss}^{(-y)} - M_{pp}^{(+y)} + M_{pp}^{(-y)}\right).
\end{align*}
With this $\delta \tilde{S}_{0} = 2 \, \mathcal{R}_{0y} \,P_{y}$ and $\delta \tilde{S}_{1} = 2 \, \mathcal{R}_{1y}\, P_{y}$. Hence, the electric field changes as
\begin{align*}
    \delta\left( |E_{s}|^{2} \right) & = \frac{\delta \tilde{S}_{0} + \delta \tilde{S}_1}{2} = (\mathcal{R}_{0y} + \mathcal{R}_{1y}) \, P_{y},\\
    \delta\left( |E_{p}|^{2} \right) &= \frac{\delta \tilde{S}_{0} - \delta \tilde{S}_1}{2} = (\mathcal{R}_{0y} - \mathcal{R}_{1y}) \, P_{y}
\end{align*}
(cf.\ Equation~\eqref{eq:es-ep}). This shows that spin reversal can result in IPE intensity changes. 

In order to corroborate the analytical calculations we performed respective photoemission calculations for W(110), described briefly in Appendix~\ref{sec:numerical}. The response tensor $\mathcal{R}(\vec{\pi})$ has been calibrated for $\vec{q}$-angles $\vartheta = \unit[45]{^{\circ}}$ and $\varphi = \unit[0]{^{\circ}}$ and a photon energy of  $\hbar \omega = \unit[9.5]{eV}$. In what follows we discuss $\vec{k}$-scans along the $\bar{\mathrm{H}}\bar{\Gamma}\bar{\mathrm{H}}$ line of the surface Brillouin zone, with a focus on energies between the Fermi energy $E_{\mathrm{F}}$ and $E_{\mathrm{F}} + \unit[4]{eV}$. 

\begin{figure}
    \centering
    \includegraphics[width=0.75\linewidth]{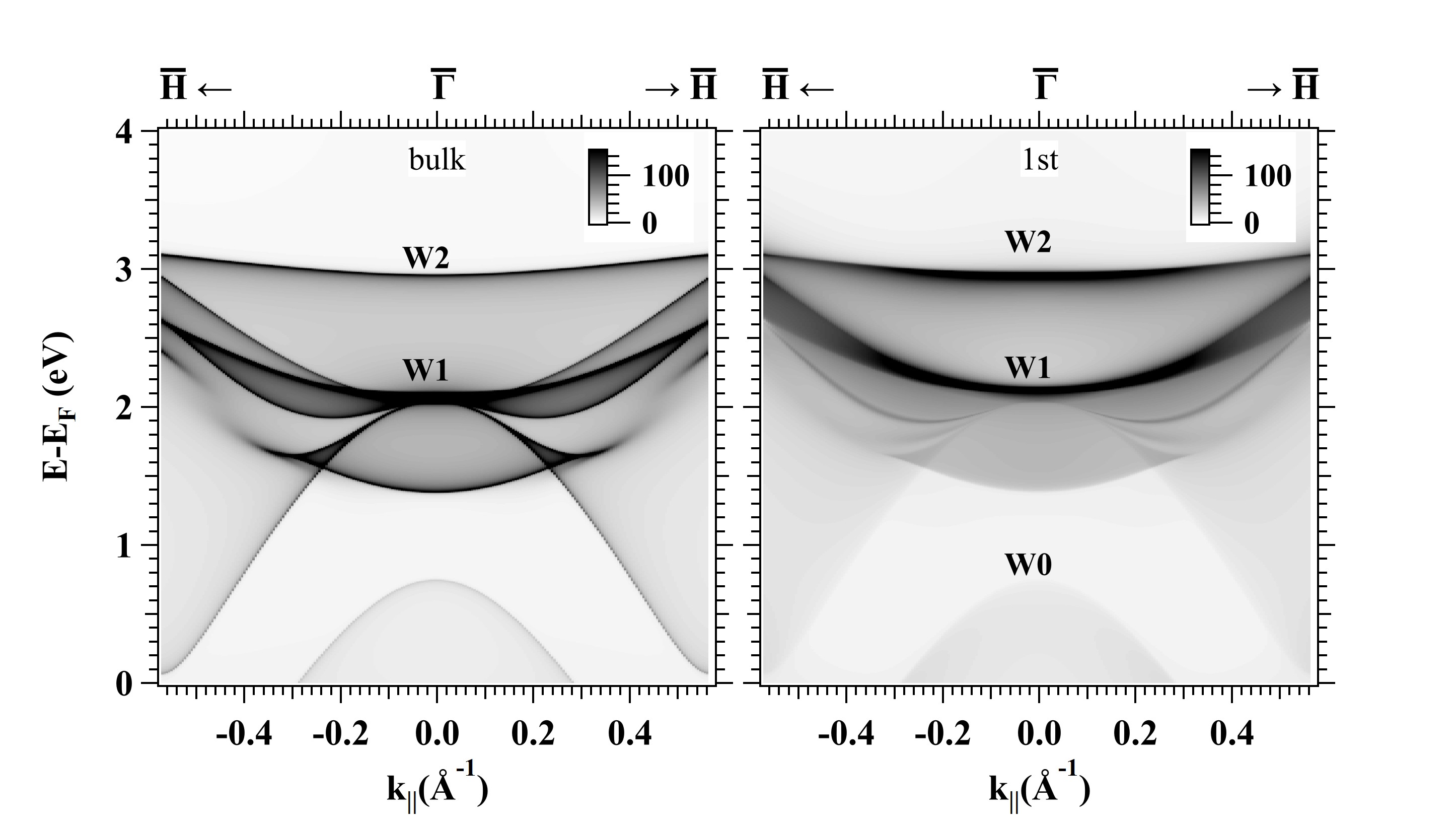}
    \caption{Spectral density $n_{l}(E, \vec{k})$ of W(110) along the $\bar{\mathrm{H}}\bar{\Gamma}\bar{\mathrm{H}}$ line of the surface Brillouin zone. The gray-scale plots show the energy- and wavevector-resolved density of states (white zero, black maximum) for a bulk layer (left) and the topmost surface layer (right) in units of $\unit{states/eV}$. $W_{0}$, $W_{1}$, and $W_{2}$ refer to bands that are discussed in the text; $E_{\mathrm{F}}$ is the Fermi energy.}
    \label{fig:SD-kx}
\end{figure}

The spectral density $n_{l}(E, \vec{k})$ (layer-, energy- and wavevector-resolved density of states; Figure~\ref{fig:SD-kx}) exhibits band edges in the bulk (dark regions in the `bulk' panel) with parabolic dispersion. These regions of large spectral density also show up in the topmost surface layer (`1st' panel) as a flat parabola at about $\unit[3]{eV}$ (labeled $W_{2}$ following Ref.~\onlinecite{Wortelen2017}) and a more dispersive parabola $W_{1}$ with a minimum at $(E = \unit[2.1]{eV}, \vec{k} = 0)$. Moreover, there is downward-dispersing parabolic band edge $W_{0}$, with a maximum at $(\unit[0.7]{eV}, 0)$. As we will see, these features play a prominent role in the SARIPES calculations.

\begin{figure}
    \centering
    \includegraphics[width=0.75\linewidth]{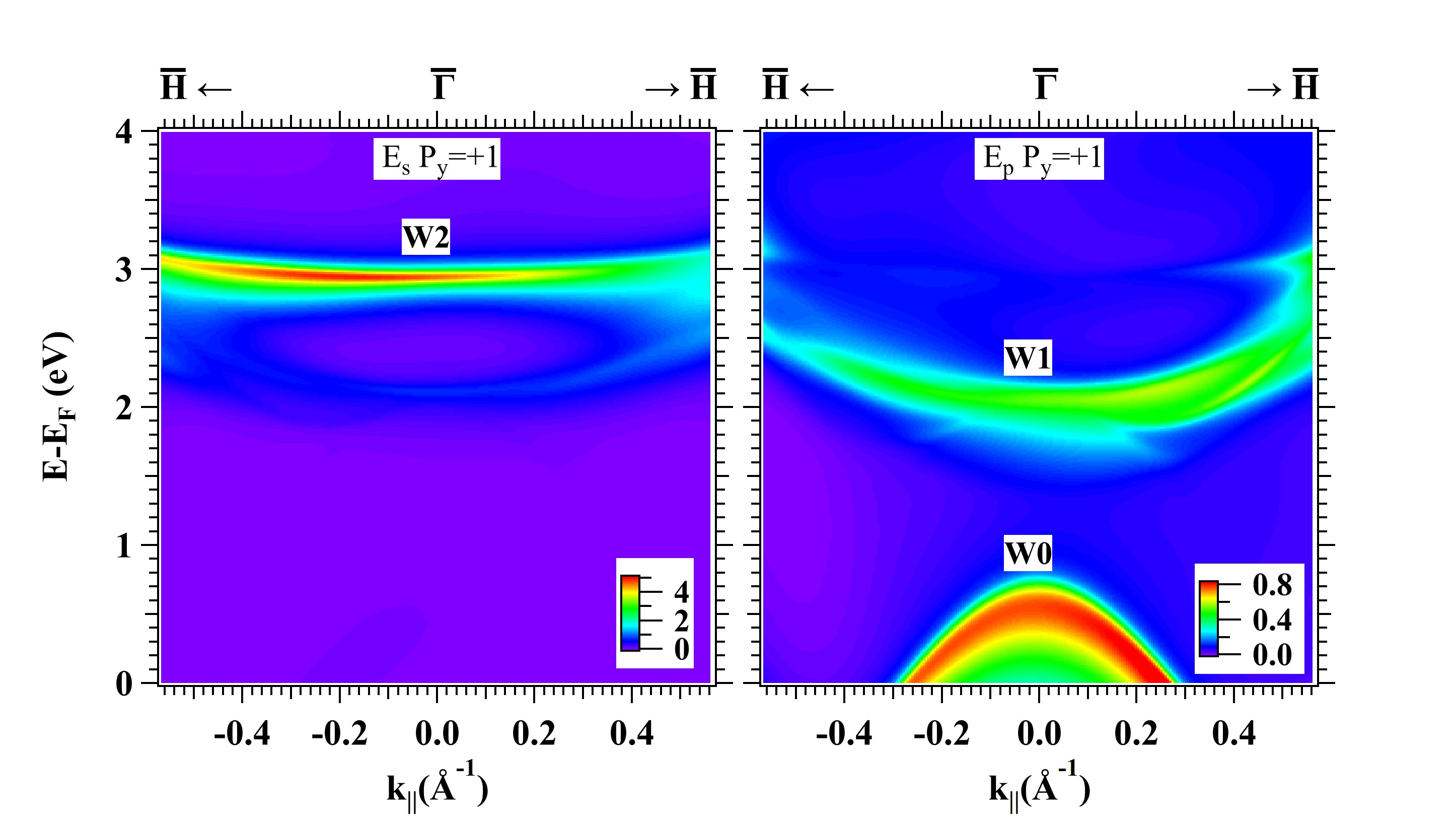}
    \caption{Emitted electric field in SARIPES from W(110). The color-scale plots show $|E_{s}|^{2}$ (left) and $|E_{p}|^{2}$ (right), respectively, with dark blue (red) minimum (maximum) intensity in arbitrary units. Both panels depict the same $(E, \vec{k})$ range as in Figure~\ref{fig:SD-kx}. The radiation propagates along $\vec{q}$ defined in the text. The incident electron is fully spin-polarized perpendicular to the scattering plane ($P_{y} = +1$).}
    \label{fig:Es-Ep-kx}
\end{figure}

For an incident fully $y$-polarized electron beam [$\vec{P} = (1, 0, +1, 0)^{\mathrm{t}}$] both s- and p-polarized radiation is emitted, in agreement with the above symmetry analysis (Figure~\ref{fig:Es-Ep-kx}; the electric fields have been calculated from the Stokes parameters $\tilde{S}_{0}$ and $\tilde{S}_{1}$ using Equation~\eqref{eq:Stokes-inverse}; $\tilde{S}_{2}$ and $\tilde{S}_{3}$ are zero). While the emitted s-polarized radiation is strongly concentrated at the upper flat band edge (left panel, labeled `$E_{s} \, P_{y} = +1$'), the p-polarized radiation is instead prominent at the other two band edges (right panel, labeled `$E_{p} \, P_{y} = +1$'), with a maximum intensity approximately one-sixth that of the s-polarized emission.

\begin{figure}
    \centering
    \includegraphics[width=\linewidth]{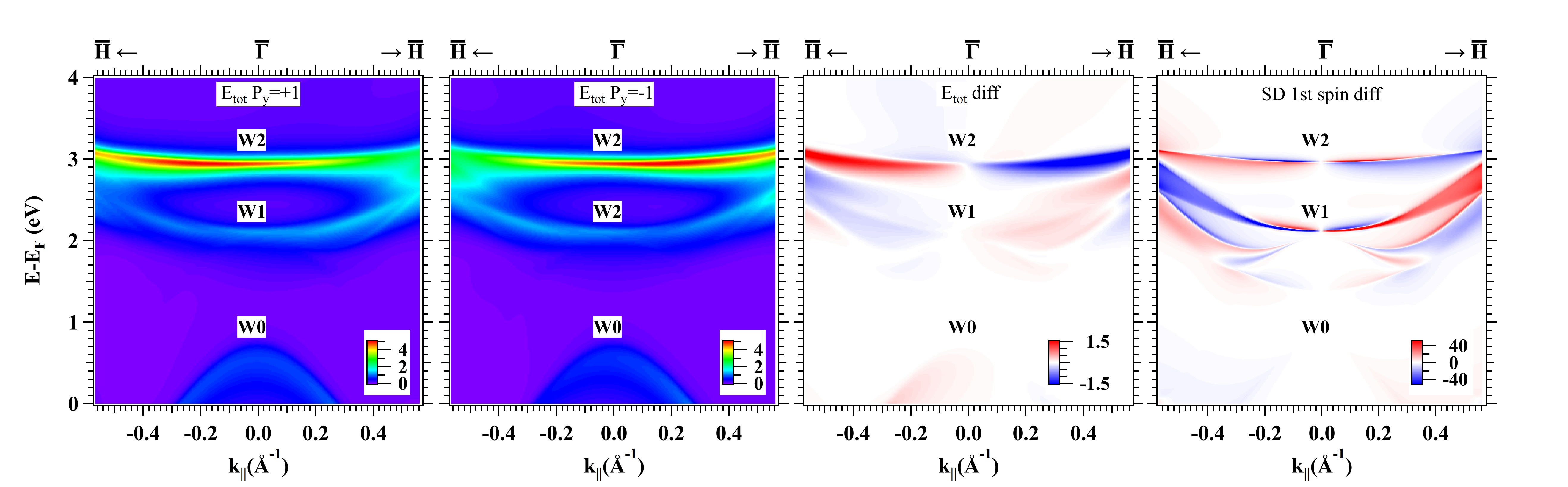}
    \caption{Spin dependence of the emitted electric field in SARIPES from W(110). The color-scale plots show $|E_{\mathrm{tot}}|^{2}$ for $P_{y} = +1$ (left panel labeled `$E_{\mathrm{tot}}\, P_{y} = +1$') and $P_{y} = -1$ (label `$E_{\mathrm{tot}}\, P_{y} = -1$'), respectively (color scale as in Figure~\ref{fig:Es-Ep-kx}). The difference of these intensities is shown in the panel labeled `$E_{\mathrm{tot}}\, \text{diff}$' (blue negative, white zero, red positive intensity). The right panel labeled `SD 1st spin diff' depicts the difference $n_{1}^{(+y)} - n_{1}^{(-y)}$ of the $y$-spin-resolved spectral densities of the topmost surface layer (in $\unit{states/eV}$). All panels depict the same $(E, \vec{k})$ range as in Figure~\ref{fig:SD-kx}. The radiation propagates along $\vec{q}$ defined in the text.}
    \label{fig:kx-compare}
\end{figure}

Both patterns exhibit a significant asymmetry with respect to $k_{x}$, which is explained by the off-normal $\vec{q}$. However, these signatures also strongly depend on the spin orientation of the electron beam, as depicted in the left panels of Figure~\ref{fig:kx-compare} for the total emitted intensity $E_{\mathrm{tot}}$ for $P_{y} = +1$ and $P_{y} = -1$. In particular, the maximum of the flat band edge $W_{2}$ at about $\unit[3]{eV}$ is shifted from negative $k_{x}$ for $P_{y} = +1$ to positive $k_{x}$ for $P_{y} = -1$ (compare the red regions). This finding suggests that spin orientation reversal translates into intensity changes in the SARIPES process. In order to corroborate this conjecture, we compare the difference of the total intensity, $|E_{\mathrm{tot}}(P_{y} = +1)|^{2} - |E_{\mathrm{tot}}(P_{y} = -1)|^{2}$, with the respective difference of the $y$-spin-resolved spectral density  $n_{1}^{(+y)} - n_{1}^{(-y)}$ of the topmost layer (layer index $l = 1$; rightmost two panels in Figure~\ref{fig:kx-compare}).

Due to Rashba-type spin–orbit interaction, electronic states in the surface layers acquire a spin polarization perpendicular to the mirror plane (along the $y$-axis; see the rightmost panel in Fig.~\ref{fig:kx-compare}), which can be probed by SARIPES\@. More precisely, when the spin polarization of the incident beam matches the spin character of the electronic state in the solid, the overlap of the corresponding spinors is maximized, leading to an enhanced signal. Conversely, a mismatch in spin orientation suppresses the signal, as the reduced spinor overlap diminishes the transition probability. Consequently, SARIPES directly reflects the spin structure of the electronic states.

We find a convincing correlation between the SARIPES intensity difference and the respective difference of the spectral density of the topmost surface layer. This correlation is most prominent for the flat band edge $W_{2}$, but also appears in other $(E, \vec{k})$ regions, say at about $\unit[2.5]{eV}$ ($W_{1}$). In addition, regions with small spin polarization (`white regions' below $\unit[2]{eV}$) coincide. While the spectral density is perfectly antisymmetric under $k_{x}$-reversal [time reversal implies $n_{l}^{(+y)}(E, \vec{k}) = n_{l}^{(-y)}(E,-\vec{k})$], a closer inspection of the respective $|E_{\mathrm{tot}}|^{2}$ data reveals tiny deviations from the strict asymmetry which are caused by the off-normal propagation $\vec{q}$ of the emitted electric radiation. Evidently, both differences should not coincide perfectly since SARIPES probes not only the topmost surface layer but also a few deeper layers and includes matrix-element effects that are taken into account by the response tensor.

We would like to point out that the described process is the time-reversed counterpart of an predicted and experimentally confirmed spin-polarization effect in conventional photoemission, in which incident linearly polarized light produces photoelectrons with spin polarization perpendicular to a mirror plane~\cite{Tamura1991,Irmer1992}.

Briefly summarizing, the above discussion shows that the proposed unified framework conforms with symmetry constraints and is able to describe successfully SARIPES intensities.

\subsubsection{Incident spin polarization within the mirror plane}
\label{sec:spin-in-plane}
If the incident electron beam is spin-polarized within the mirror plane ($P_{x}$ or $P_{z}$ nonzero, $P_{y} = 0$), Equation~\eqref{eq:S-of-P} gives
\begin{align*}
    \tilde{S}_{0} & = \mathcal{R}_{00}, \\
    \tilde{S}_{1} & = \mathcal{R}_{10}, \\
    \tilde{S}_{2} & = \mathcal{R}_{2x} \, P_{x} + \mathcal{R}_{2z} \, P_{z},\\
    \tilde{S}_{3} & = \mathcal{R}_{3x} \, P_{x} + \mathcal{R}_{3z} \, P_{z}.
\end{align*}
The emitted electric field is then parameterized as 
\begin{subequations}
    \label{eq:S-of-P-inplane}
\begin{align}
    |E_{s}|^{2} & = \frac{\mathcal{R}_{00} + \mathcal{R}_{10}}{2}, \\
    |E_{p}|^{2} & = \frac{\mathcal{R}_{00} - \mathcal{R}_{10}}{2}, \\
    \tan\delta & = \frac{\mathcal{R}_{3x} \, P_{x} + \mathcal{R}_{3z} \, P_{z}}{\mathcal{R}_{2x} \, P_{x} + \mathcal{R}_{2z} \, P_{z}}.
\end{align}    
\end{subequations}
This already tells us that only the phase difference $\delta$ between the $s$- and $p$-polarized field components depends on the spin polarization, while the field's magnitude (intensity) does not.

We now consider a reversal of the spin polarization, $P_{x} \to -P_{x}$ and $P_{z} \to -P_{z}$. According to Table~\ref{tab:symmetry-2mm}, such a reversal is accompanied by sign changes for $S_{2}$ and $S_{3}$, but not for $S_{0}$ and $S_{1}$, which is fully in line with the preceding expressions~\eqref{eq:S-of-P-inplane}. In short, the spin reversal leaves the intensity of the emitted elliptically polarized radiation invariant~\footnote{The two setups $(P_{x}, P_{z})$ and $(-P_{x}, -P_{z})$ are connected by the mirror operation. In the sense emphasized by Wilczek, symmetry can be understood as `change without change,' i.\,e., invariance under a specified transformation~\cite{Wilczek20215}, implying that the total intensity of the two setups is identical. In contrast, the two setups discussed in Section~\ref{sec:spin-perp-plane} are not related by a symmetry operation and, consequently, their total intensities are different.}. The present scenario can be viewed as the time-reversed counterpart of optical orientation.

For the setup described in Section~\ref{sec:spin-perp-plane}, spin information was encoded in the SARIPES intensity change upon reversal of the spin polarization of the incident electron beam. In the present case, no such intensity change is observed upon spin reversal, suggesting that the degree of circular polarization $S_{3} / S_{0}$ may instead be linked to the spin polarization of the sample's electronic states.

\begin{figure}
    \centering
    \includegraphics[width = \linewidth]{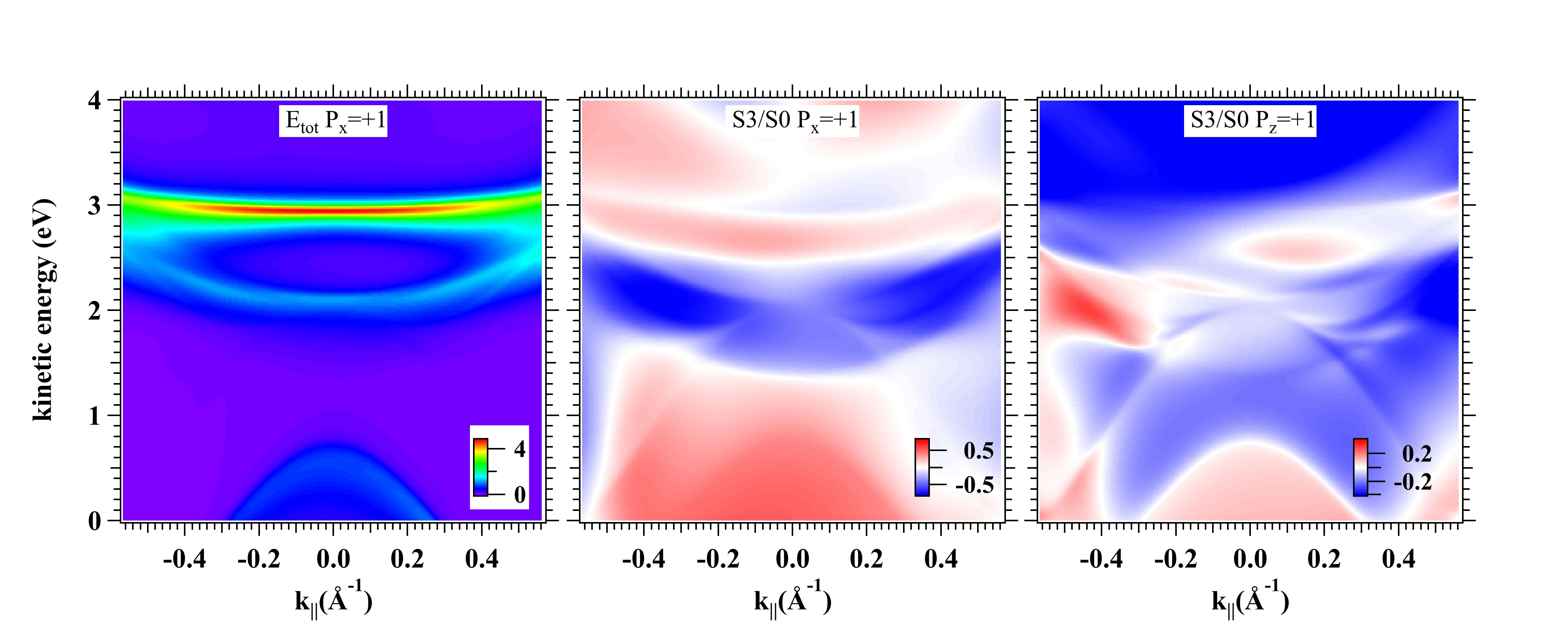}
     \caption{Spin dependence of the emitted electric field in SARIPES from W(110). The left color-scale plot shows $|E_{\mathrm{tot}}|^{2}$ for an $x$-polarized incident electron beam ($P_{x} = +1$; labeled `$E_{\mathrm{tot}} \, P_{x} = +1$'; color scale as in Figure~\ref{fig:Es-Ep-kx}). The respective degree of helicity of the emitted radiation is depicted in the central panel labeled `$S3/S0 \, P_{x} = +1$' (blue negative, white zero, red positive values). The right panel labeled `$S3/S0 \, P_{z} = +1$' displays the helicity degree for a $z$-polarized incident beam ($P_{z} = +1$). All panels cover the same $(E, \vec{k})$ range as in Figure~\ref{fig:SD-kx}. The radiation propagates along $\vec{q}$ defined in the text.}
    \label{fig:tungsten-pxz}
\end{figure}

In order to test this conjecture, we computed SARIPES spectra for incident electrons fully spin-polarized along the $x$- and $z$-directions (Figure~\ref{fig:tungsten-pxz}). In both cases, the strongest SARIPES signal originates from state $W_{2}$ (left panel), and the helicity reverses sign upon reversal of the incident-spin polarization (e.g., $P_{x} = +1 \rightarrow -1$; not shown), in agreement with the symmetry constraints. For the $x$-polarized beam, the $(E,\vec{k})$ distribution of the degree of helicity exhibits a partial resemblance to the spectral density (center panel), whereas for the $z$-polarized beam such correspondence becomes even less pronounced (right panel). The absence of a clear correlation with the electronic structure indicates that, in the present geometry, the degree of circular polarization is governed predominantly by the photoemission matrix elements rather than by the spin-dependent spectral density.

In summary, the setup discussed in this Section is less suitable for retrieving the spin polarization of electronic states within a mirror plane than the geometry considered in Section~\ref{sec:spin-perp-plane}. As a rule of thumb, spin retrieval is most effective when the probed spin polarization and the incident spin polarization are aligned.

\section{Concluding remarks}
We have established a unified theoretical framework for spin- and angle-resolved photoemission and its inverse, centered on a response tensor that encodes the reciprocity between SARPES and SARIPES\@. Symmetry constraints reduce the number of independent components, leading to explicit expressions for spin-dependent transitions and intensity modulations. Moreover, polarization tomography based on SARPES calculations enables the complete reconstruction of the underlying response functions (Appendix~\ref{sec:strategy-response-tensor}). This symmetry-adapted formalism provides a common quantitative description of spin-dependent optical transitions at surfaces, allowing SARIPES intensities to be predicted directly from SARPES calculations. 

An important direction for future work is the extension of the present formalism to magnetic systems. In such cases, time-reversal symmetry is broken and additional spin-dependent terms may appear in the response tensor. Incorporating magnetic symmetry groups and magnetization-dependent selection rules will allow the framework to describe spin-polarized transitions in ferromagnetic and antiferromagnetic materials, opening the way to a systematic analysis of magnetic order and spin textures through combined SARPES and SARIPES analyses.

\section*{Acknowledgments}
We would like to thank Markus Donath (Münster) for bringing this topic to our attention and Roland Feder (Duisburg-Essen) for valuable discussions. The authors used artificial intelligence to assist with language editing and text refinement and take full responsibility for the content.

\appendix

\section{Motivation of the response tensor}
\label{sec:derivation-response-tensor}
The spin-resolved SARPES intensity~\eqref{eq:intensity} expressed in the $(s,p)$ polarization basis of the incident radiation can be rewritten in terms of the Stokes parameters $S_{i}$, Equation~\eqref{eq:intensity-Stokes-basis}. Comparison of both expressions yields the coefficients $a_{i}^{(\pm \sigma)}$ defined in Equation~\eqref{eq:coeffs-Stokes-basis}. With the $4$-vector $\vec{P} = (P_{0}, P_{x}, P_{y}, P_{z})^{\mathrm{t}}$ ($P_{0} = 1$) the intensities can be written as the bilinear form~\eqref{eq:bilinear}. Each element $\mathcal{R}_{\alpha\beta}$ is a linear combination of the response functions $M_{ij}^{(\sigma)}$. Comparison of both forms, Equations~\eqref{eq:intensity-Stokes-basis} and~\eqref{eq:bilinear}, gives the following relations (spin axis $\sigma = x, y, z$ or $1, 2, 3$).
\begin{description}
    \item[Related to $S_{0}$] (unpolarized component)
\begin{align*}
    \mathcal{R}_{0 0} &= \frac{1}{2} \left(M_{ss}^{(+x)} + M_{ss}^{(-x)} + M_{pp}^{(+x)} + M_{pp}^{(-x)}\right),\\
    \mathcal{R}_{0 \sigma} &= \frac{1}{2} \left(M_{ss}^{(+\sigma)} - M_{ss}^{(-\sigma)} + M_{pp}^{(+\sigma)} - M_{pp}^{(-\sigma)}\right).
    \end{align*}

    \item[Related to $S_{1}$] (linear $s$, $p$)
\begin{align*}
    \mathcal{R}_{1 0} &= \frac{1}{2} \left(M_{ss}^{(+x)} + M_{ss}^{(-x)} - M_{pp}^{(+x)} - M_{pp}^{(-x)}\right),\\
    \mathcal{R}_{1 \sigma} &= \frac{1}{2} \left(M_{ss}^{(+\sigma)} - M_{ss}^{(-\sigma)} - M_{pp}^{(+\sigma)} + M_{pp}^{(-\sigma)}\right).
\end{align*}

\item[Related to $S_{2}$]  (linear $45^{\circ}$)
\begin{align*}
    \mathcal{R}_{2 0} &= \mathrm{Re}\,M_{sp}^{(+x)} + \mathrm{Re}\,M_{sp}^{(-x)},\\
    \mathcal{R}_{2 \sigma} &= \mathrm{Re}\,M_{sp}^{(+\sigma)} - \mathrm{Re}\,M_{sp}^{(-\sigma)}.
\end{align*}

\item[Related to $S_{3}$] (circular $45^{\circ}$)
\begin{align*}
    \mathcal{R}_{3 0} &= -\mathrm{Im}\,M_{sp}^{(+x)} - \mathrm{Im}\,M_{sp}^{(-x)},\\
    \mathcal{R}_{3 \sigma} &= -\mathrm{Im}\,M_{sp}^{(+\sigma)} + \mathrm{Im}\,M_{sp}^{(-\sigma)}.
    \end{align*}
\end{description}
Since the elements $\mathcal{R}_{\alpha 0}$ comprise sums over opposite spin orientations (here: $\sigma = +x$ and $\sigma = -x$), they do not depend of the spin axis [$x$, $y$ or $z$; compare Equation~\eqref{eq:Pdef}].

Let us consider a setup that is sensitive to $p$-polarized light in the $xz$-plane and to $y$-polarized electrons; hence
\begin{align*}
    \vec{S} & = \begin{pmatrix}
        1 \\ -1 \\ 0 \\ 0
    \end{pmatrix} |E|^{2},
    \quad
    \vec{P} = \begin{pmatrix}
        1 \\ 0 \\ P_{y} \\ 0
    \end{pmatrix}.
\end{align*}
Evaluating the bilinear expression $I =  \vec{S}^{\mathrm{t}} \mathcal{R} \vec{P}$ yields explicitly
\begin{align*}
    I & = 
    \left[ 
        \left(M_{pp}^{(+y)} + M_{pp}^{(-y)}\right) + \left(M_{pp}^{(+y)} - M_{pp}^{(-y)}\right) P_{y} 
    \right] |E|^{2}.
\end{align*}
The intensity decomposes into a spin-averaged contribution (term in the first parentheses) and a spin-dependent contribution (term in the second parentheses).

As another example, we consider a setup involving left- or right-handed circularly polarized light in the $xz$-mirror plane and $z$-polarized electrons,
\begin{align*}
    \vec{S} & = \begin{pmatrix}
        1 \\ 0 \\ 0 \\ \pm 1
    \end{pmatrix} |E|^{2},
    \quad
    \vec{P} = \begin{pmatrix}
        1 \\ 0 \\ 0 \\ P_{z} 
    \end{pmatrix}.
\end{align*}
The bilinear expression gives, exploiting the constraints from the mirror operation [Equation~\eqref{eq:R-in-mirror-plane}] $I = \left( \mathcal{R}_{00} \pm \mathcal{R}_{3z} P_{z}\right) |E|^{2}$ or
\begin{align*}
    I = 
    \left( \frac{1}{2} \left(M_{ss}^{(+z)} + M_{ss}^{(-z)} + M_{pp}^{(+z)} + M_{pp}^{(-z)}\right) \mp (\mathrm{Im}\,M_{sp}^{(+z)} - \mathrm{Im}\,M_{sp}^{(-z)}) P_{z}\right) |E|^{2}.
\end{align*}
As in the previous example, the intensity decomposes into spin-averaged and spin-dependent contributions. Moreover, simultaneously reversing the light's helicity and the spin polarization keeps the intensity unchanged, as it has to be for setups related by a symmetry operation (Table~\ref{tab:symmetry-2mm}).

It appears natural to separate unpolarized (0) and polarized (1--3) components,
\begin{align*}
    \mathcal{R} & =
    \begin{pmatrix}
    R_{00} & \vec{R}_{0v}^{\mathrm{t}} \\
    \vec{R}_{v0} & \mat{R}_{vv}
    \end{pmatrix}.
\end{align*}
Each block corresponds to a distinct physical effect (subscripts read from left to right):
\begin{itemize}
    \item $R_{00}$: ordinary intensity,
    
    \item $\vec{R}_{0v}^{\mathrm{t}}$: spin generation from unpolarized light,
    
    \item $\vec{R}_{v0}$: dichroism effects (varying the light's polarization alters the spin-averaged electron current),
    
    \item $\mat{R}_{vv}$: photon polarization--spin coupling, further decomposed into isotropic, antisymmetric, and symmetric traceless components,
    \begin{align*}
    \mat{R}_{vv} = 
    \underbrace{\frac{1}{3} \tr (\mat{R}_{vv})\, \mat{1}_{3 \times 3}}_{\text{isotropic scalar}} 
    + \underbrace{\frac{1}{2} (\mat{R}_{vv}-\mat{R}_{vv}^{\mathrm{t}})}_{\text{antisymmetric}} 
    + \underbrace{\frac{1}{2} (\mat{R}_{vv}+\mat{R}_{vv}^{\mathrm{t}}) - \frac{1}{3}\tr(\mat{R}_{vv}) \mat{1}_{3 \times 3}}_{\text{symmetric traceless}}
\end{align*}
\end{itemize}
This decomposition shows that spin- and polarization-resolved photoemission provides a complete characterization of the polarization transfer properties of the photoemission process, analogous to quantum tomography~\cite{Mauro2003}.

In order to provide more context, we note that the polarization degrees of freedom of both photons and the electrons are described by an underlying $\mathrm{SU}(2)$ structure, such that the parameters $S_{\alpha}$ and $P_{\beta}$ appear as expansion coefficients in a common algebra~\footnote{The Stokes parameters can be interpreted as expectation values of Pauli matrices with respect to the polarization (coherency) matrix, establishing a direct correspondence to the density-matrix formalism of a two-level quantum system (see, e.g., Refs.~\cite{Mandel1995,Loudon2000,Luis2004}).}. Within this framework, the response tensor $\mathcal{R}$ represents a bilinear form in the direct product space of two $\mathrm{SU}(2)$ algebras. This makes explicit that the photoemission process couples photon polarization and electron spin on equal footing, with $\mathcal{R}$ acting as a transfer operator between the two spaces. The above decomposition of $\mathcal{R}$ into scalar, vector, and tensor contributions corresponds to the irreducible components under $\mathrm{SU}(2)$ rotations, providing a natural classification of the symmetry-allowed polarization transfer channels.

\section{Construction of the response tensor}
\label{sec:strategy-response-tensor}
The response tensor $\mathcal{R}(\vec{\pi})$ encodes the complete spin-dependent SARPES and SARIPES response, as it mediates the coupling between photon polarization and electron spin degrees of freedom. For a given setup
$\vec{\pi}$, the bilinear form $I(\vec{\pi}; \vec S, \vec P)$, Equation~\eqref{eq:bilinear}, provides an equivalent representation of this
response. In particular, $\mathcal{R}$ is encoded in generating
form, in the sense that its elements can be recovered as
\begin{align*}
    \mathcal{R}_{\alpha \beta}
    = \frac{\partial^2 I}{\partial S_{\alpha} \, \partial P_{\beta}}.
\end{align*}
Thus, $I$ acts as a generating function for the response coefficients $\mathcal{R}_{\alpha \beta}$, and the full tensor can be reconstructed by systematically probing the bilinear form with independent variations of $\vec{S}$ and $\vec{P}$. For this polarization tomography, we suggest the following strategy.
\begin{enumerate}
\item Compute SARPES intensities
\begin{align*}
    I^{(\pm \sigma)}(E_{s}, E_{p}) & = M_{ss}^{(\pm \sigma)} |E_{s}|^{2} + M_{pp}^{(\pm \sigma)} |E_{p}|^{2} + 2\,\mathrm{Re}(E_{s}  M_{sp}^{(\pm \sigma)} E_{p}^{\star}), 
\end{align*}
[Equation~\eqref{eq:intensity} with $M_{ps}^{(\pm \sigma)} = (M_{sp}^{(\pm \sigma)})^{\star}$; $\sigma \in \{ x, y, z \}$] for six photon polarization states constructed from $(E_{s}, E_{p})$ with equal amplitudes (right column):
\begin{align*}
I_{s}^{(\pm \sigma)} & = M_{ss}^{(\pm \sigma)}, & (1, 0)
\\
I_{p}^{(\pm \sigma)} & = M_{pp}^{(\pm \sigma)}, & (0, 1)
\\
I_{+45}^{(\pm \sigma)} & = \frac{1}{2} \left( M_{ss}^{(\pm \sigma)} + M_{pp}^{(\pm \sigma)} \right) + \Re\, M_{sp}^{(\pm \sigma)}, & \frac{\sqrt{2}}{2} (1, 1 )
\\
I_{-45}^{(\pm \sigma)} & = \frac{1}{2} \left( M_{ss}^{(\pm \sigma)} + M_{pp}^{(\pm \sigma)} \right) - \Re\, M_{sp}^{\pm \sigma)}, & \frac{\sqrt{2}}{2} (1, -1) 
\\
I_{+c}^{(\pm \sigma)} & = \frac{1}{2} \left( M_{ss}^{(\pm \sigma)} + M_{pp}^{(\pm \sigma)} \right) - \Im\, M_{sp}^{(\pm \sigma)}, & \frac{\sqrt{2}}{2} (1, \mathrm{i})
\\
I_{-c}^{(\pm \sigma)} & = \frac{1}{2} \left( M_{ss}^{(\pm \sigma)} + M_{pp}^{(\pm \sigma)} \right) + \Im\, M_{sp}^{(\pm \sigma)}. &  \frac{\sqrt{2}}{2} (1, -\mathrm{i}) .
\end{align*}

\item Optional: normalize all spin-resolved intensities, for example
\begin{align*}
    I_{s}^{(\pm \sigma)} & \to \frac{I_{s}^{(\pm \sigma)}}{I_{s}^{(+\sigma)} +I_{s}^{(-\sigma)}}
    = \frac{1}{2} \left( 1 \pm P_{s}^{(\sigma)} \right).
\end{align*}

\item  Solve for the response functions in terms of the computed intensities:
\begin{align*}
M_{ss}^{(\pm \sigma)} & = I_{s}^{(\pm \sigma)}, 
\\
M_{pp}^{(\pm \sigma)} &= I_{p}^{(\pm \sigma)}, 
\\
\mathrm{Re}\, M_{sp}^{(\pm \sigma)} & = \frac{1}{2} \left( I_{+45}^{(\pm \sigma)} - I_{-45}^{(\pm \sigma)} \right),
\\
\mathrm{Im}\, M_{sp}^{(\pm \sigma)} & = \frac{1}{2} \left( I_{-c}^{(\pm \sigma)} - I_{+c}^{(\pm \sigma)} \right).
\end{align*}

\item Compute the tensor elements $\mathcal{R}_{\alpha \beta}$ that are explicitly given in Appendix~\ref{sec:derivation-response-tensor}.
\end{enumerate}

\section{Details of the numerical calculations}
\label{sec:numerical}
The foundation of the numerical results reported in this Paper are density-functional calculations for W(110), as reported in Ref.~\onlinecite{Wortelen2017}. Using the self-consistent potential as input, we calculated the electronic structure, spin- and layer-resolved spectral densities, as well as spin-and angle-resolved photoemission spectra within the one-step model using our program package \textsc{omni}~\cite{Henk2018}. This codes relies on the spin-polarized relativistic layer KKR (Korringa-Kohn-Rostoker) method (see e.g., Ref.~\onlinecite{Henk2002} and references therein). By solving the Dirac equation for the single-site scattering problem, spin-orbit coupling is naturally included, which is essential for spin-resolved spectroscopies from nonmagnetic samples~\cite{Feder1996}.

We focus on an energy window relevant for inverse photoemission, that is $[E_{\mathrm{F}}, E_{\mathrm{F}} + \unit[4]{eV}]$ and a photon energy of $\unit[9.5]{eV}$. Since $\textsc{omni}$ is based on Green functions, lifetime effects have to be included by a complex self-energy~\cite{Feibelman1974,Tusche2018}, whose imaginary part was taken as $\unit[50]{meV}$ in order to approximate well the Fermi's golden rule--expression for photoemission (for the spectral-density calculations  $\unit[5]{meV}$ were chosen). Such small a lifetime broadening required extensive convergence tests, concerning for example the number of layers, maximum angular momentum $l_{\mathrm{max}}$ in the single-site expansion, etc. Moreover, focusing on proof-of-principle calculations in this Paper, we set the dielectric constant $\epsilon = 1$.

Focusing on emission within a mirror plane (Section~\ref{sec:example}), we consider the $\bar{\mathrm{H}}\bar{\Gamma}\bar{\mathrm{H}}$ line in the two-dimensional Brillouin zone. The resulting spin-resolved spectral densities agree well with those reported in Ref.~\onlinecite{Wortelen2017}. Respective SARPES calculations were then performed to construct the response tensor $\mathcal{R}$ for each $(E, \vec{k})$, following the strategy described in Appendix~\ref{sec:strategy-response-tensor}. The computed tensor elements obey the symmetry constraints and reproduce the input/output SARPES data with a relative error better than $10^{-6}$. Physical consistency was checked by the positivity of both $\tilde{\vec{S}}$ and $\tilde{\vec{P}}$, putting the SARIPES results reported in Section~\ref{sec:example} on a firm basis.

\section{Supporting material}
\label{sec:supporting}
In this Section we provide additional information concerning the reciprocity of SARPES and SARIPES as well as on the projection concept discussed in Section~\ref{sec:response-tensor}.

\begin{figure*}
    \centering
    \includegraphics[width = \linewidth]{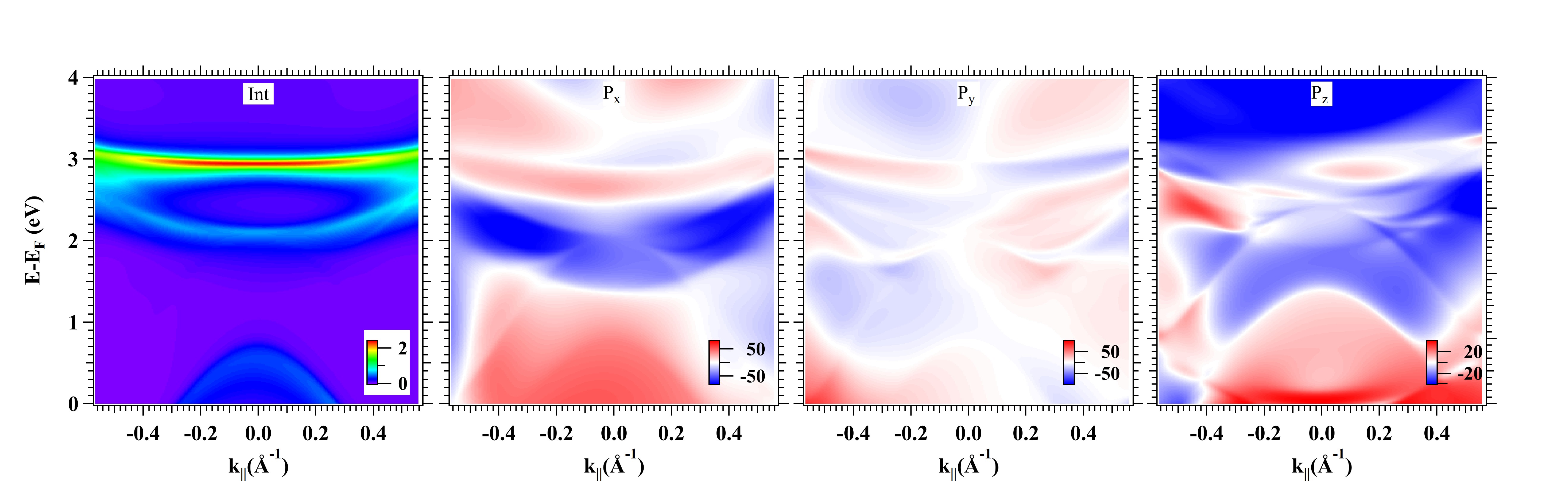}
    \caption{Spin- and angle-resolved PE spectra for circular polarized light. The color scales show the total intensity and the spin-resolved intensities (from left to right, as indicated).}
    \label{fig:tungsten-xyz-PE}
\end{figure*}

The reciprocity between SARPES and SARIPES is illustrated by comparing the spin-resolved PE spectra obtained with circularly polarized light (Figure~\ref{fig:tungsten-xyz-PE}) with their IPE counterparts (Figure~\ref{fig:tungsten-pxz}). In particular, the $P_x$ and $P_z$ components in Figure~\ref{fig:tungsten-xyz-PE} exhibit features that closely resemble those observed in the corresponding $S3/S0$ data in Figure~\ref{fig:tungsten-pxz}.

\begin{figure*}
    \centering
    \includegraphics[width = \linewidth]{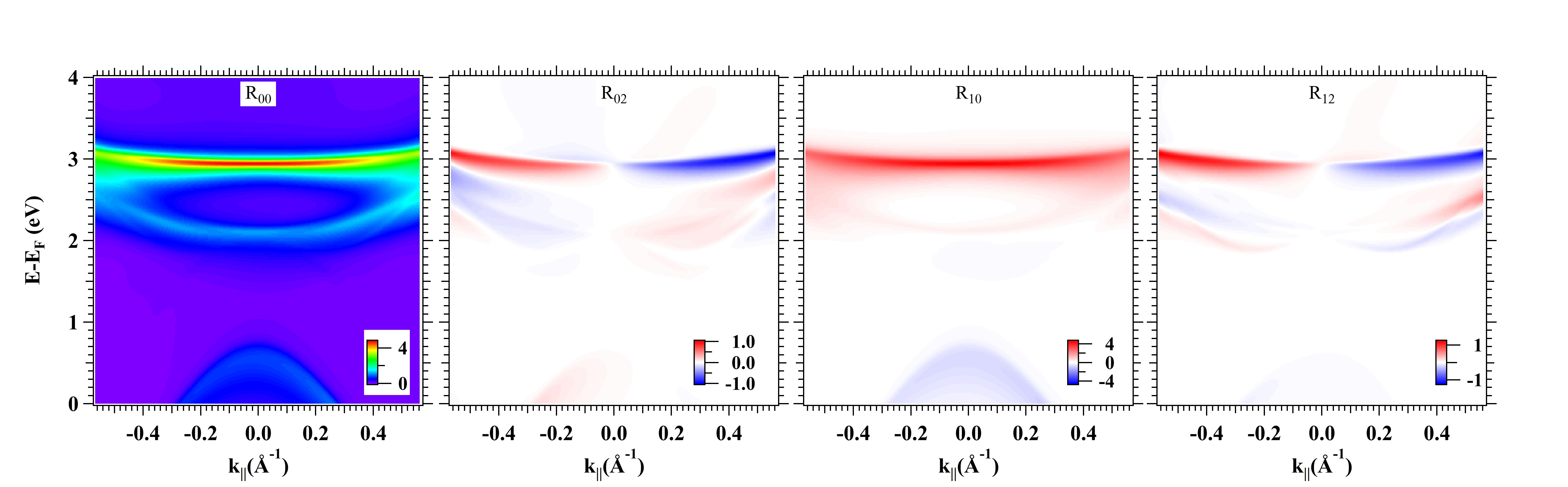}
    \caption{Elements of the spin-Stokes response tensor relevant for a $y$-polarized incident electron beam in the SARIPES context (as indicated). See  Section~\ref{sec:spin-perp-plane}.}
    \label{fig:rs-for-py}
\end{figure*}

In Section~\ref{sec:response-tensor}, we argued that the contractions in both the SARPES and SARIPES contexts can be interpreted as projections onto a detector setting. This interpretation is further supported by inspecting the elements of the response tensor. In Figure~\ref{fig:rs-for-py}, we show the elements relevant for a $y$-polarized incident electron beam (see Section~\ref{sec:spin-perp-plane}). These elements are combined through the SARIPES contraction, Equation~\eqref{eq:Stokes-inverse}, to yield the effective emitted electromagnetic radiation and therefore contain the relevant information. The components $P_{\beta}$ of the polarization vector then act as weighting factors in the corresponding contraction,
\begin{align*}
\tilde{S}{\alpha}(\vec{\pi}; \vec{P})
= \sum{\beta = 0}^{3} \mathcal{R}{\alpha\beta}(\vec{\pi}) , P{\beta},
\quad \alpha = 0, \ldots, 3,
\end{align*}

\bibliographystyle{apsrev4-2}
\bibliography{literature}

\end{document}